\documentstyle[twoside,fleqn,espcrc2,epsfig,psfig]{article}

\newcommand{\AmS}{{\protect\the\textfont2
  A\kern-.1667em\lower.5ex\hbox{M}\kern-.125emS}}
\def\NT{N_\tau}
\def\nt{\ifmmode\NT\else$\NT$\fi}
\def\NS{N_\sigma}
\def\ns{\ifmmode\NS\else$\NS$\fi}

\def\PR{{ Phys.\ Rev.\ }}

\def\PL{{ Phys.\ Lett.\ }}
\def\NP{{ Nucl.\ Phys.\ }}
\def\OP{{\langle M \rangle }}
\def\OPQ{{\langle M^2 \rangle }}
\def\MO{{\langle |M| \rangle }}

\title{Corrections to Scaling and Critical Amplitudes \\
         in SU(2) Lattice Gauge Theory    
         \thanks{We acknowledge
financial support of the European Commission under the
TMR-Programme ERBFMRX-CT97-0122}
}

\author{ J. Engels and T. Scheideler\address{Fakult\"at f\"ur Physik, 
Universit\"at Bielefeld,
D-33615 Bielefeld, Germany}} %
        
\begin{document}

\begin{abstract}
We calculate the critical amplitudes of the Polyakov loop and its
susceptibility at the deconfinement transition of $SU(2)$ gauge theory.
To this end we carefully study the corrections to the scaling functions
of the observables coming from irrelevant exponents. As a guiding line
for determining the critical amplitudes we use envelope equations derived
from the finite size scaling formulae for the observables. The equations are
then evaluated with new high precision data obtained on $\ns^3\times4$
lattices for $\ns=12,18,26$ and 36. We find different correction-to-scaling 
behaviours above and below the transition. Our result for the 
universal ratio of the susceptibility amplitudes
is $C_+/C_-=4.72(11)$ and agrees perfectly with a recent measurement for 
the $3d$ Ising model.
\end{abstract}

\maketitle

\section{ INTRODUCTION}

          In systems which exhibit a second order transition in the 
          thermodynamic limit ($V\rightarrow\infty$) the critical 
          observables behave as           
          \begin{equation}
            O_{\infty} = a_0|t|^{-\rho} {\rm ~for~} |t|\rightarrow 0.
          \label{limit}
          \end{equation}
          The coefficient $a_0$ is the critical amplitude, 
          $\rho$ the
          critical exponent of the observable $O$. The variable $t$ is
          the reduced temperature $t=(T-T_c)/T_c$. In particular, the 
          correlation length $\xi$, the magnetization or order parameter
          $\OP$ and its susceptibility $\chi$ behave for zero external 
          magnetic field $H$ close to the transition as follows
           \begin{eqnarray}
            \xi &=& f_{\pm}|t|^{-\nu}~,  \\
            \OP &=& B(-t)^{\beta} {\rm ~for~} t < 0~, \\
            \chi&=& C_{\pm}|t|^{-\gamma}~. 
           \end{eqnarray}
          The index on the amplitude refers to the symmetric (+)
          or to the broken phase ($-$).

         \indent Certain ratios of critical amplitudes, for example
     $C_+/C_-$ or $f_+/f_-$
          are  universal \cite{Cramp}, like the critical exponents, that is
          for systems of the same universality class
          they are equal. The $3d$ Ising model and $SU(2)$ gauge theory in
          (3+1) dimensions are in one class. In the Ising model such ratios
          have been calculated \cite{Case97} using very large lattices.

          The aim of this paper is the calculation of the critical amplitudes 
          $B,C_+$ and $C_-$ for $SU(2)$ with data from moderate size lattices
          using finite size scaling techniques.

\section{FINITE SIZE SCALING}
          In volumes $V$ with a characteristic length scale  
          $L=V^{1/d}$ finite size scaling theory predicts the following
          scaling form for small $|t|$
          \begin{equation}
          O(t,L) = L^{\rho / \nu} \cdot Q(tL^{1/{\nu}},L^{-\omega})~.
          \end {equation}
          Here, $H=0$ and only the largest irrelevant exponent $\lambda=-\omega$
          was taken into account.

          The functions $O(t,L)$ build a family of curves, parametrized by $L$.
          We calculate the envelope function to this family. At
          least the leading term in $t$ should coincide with the limiting form 
          (\ref{limit}). 

          The scaling function $Q(x,y)$ depends on the scaled reduced temperature
          $x=tL^{1/\nu}$ and the correction-to-scaling variable
          $y=L^{-\omega}$. We assume a linear dependence of $Q$ on
          $y$
          \begin{equation}
          Q(x,y)=Q_0(x)+yQ_1(x)~.        
          \end{equation}
\noindent  The envelope function $O_e$ is then
          \begin{equation}     
O_e = ({t \over x_0})^{-\rho}\left\{ Q_0 + ({t \over x_0})^{\omega\nu}
Q_1 + O(|t|^{2\omega\nu}) \right\}.
          \label{envel}
          \end{equation}
           
\noindent Here the $Q_i$ have to be taken at $x_0$, where $x_0$ is
        a zero of the function $F_O$
        \begin{equation}
        F_O=\rho Q_0(x)+ x Q_0^{\prime}(x)~.
        \end{equation}
\noindent        Comparing the eqs. (\ref{limit}) and (\ref{envel}) we find
        \begin{equation}
        a_0 = |x_0|^{\rho} Q_0(x_0)~.
        \end{equation}
\noindent This result is not a surprise. Suppose, there are no corrections 
to scaling, so that         
          \begin{equation}
          O(t,L) = L^{\rho/\nu} Q_0(x) = |t|^{-\rho} |x|^{\rho} Q_0(x)~,
          \end{equation} 
          and in the thermodynamic limit
          \begin{equation}
         O_{\infty} = |t|^{-\rho} \lim \limits_{ x \to \infty} |x|^{\rho} Q_0(x)~.
          \end{equation}
           We consider therefore the amplitude function $A_0$
          \begin{equation}
          A_0(x) = |x|^{\rho} Q_0(x)~,
          \end{equation}
\noindent     and its approach to asymptopia. Its derivative is given by
          \begin{equation}
          A_0^{\prime}(x) = {\rm sign}(x)|x|^{\rho-1} F_O(x)~.
          \end{equation}
          The function $F_O$ is zero at the point $x_0$ and also if 
          $Q_0(x)\sim |x|^{-\rho}$, that is 
          when $Q_0$ has reached its asymptotic form. Then $A_0(x)$
          attains an extreme value, the critical point amplitude $a_0$. 

          From the above it is clear how to proceed:         
          We measure the observable $O$ for several fixed $L$ and 
          calculate the scaling function $Q=L^{-\rho/\nu}O$. Then
          we determine $Q_0$ and $Q_1$ by a linear fit in $y$ 
          of $Q$ at fixed $x$ and               
          control the approach to the asymptotic scaling form 
          by calculating $F_O(x)$.   
          When $F_O$ stays essentially zero, we evaluate $A_0(x)=a_0$.
          In addition the correction-to-scaling amplitude may be estimated from
          \begin{eqnarray}
          a_1 = |x|^{-\omega\nu}Q_1(x)/Q_0(x).
          \end{eqnarray}

\section{DATA FOR $SU(2)$ GAUGE THEORY}
          We use the standard Wilson action and work on $N_{\sigma}^3\times4$
          lattices. Taking the lattice spacing $a=1$ makes
          $N_{\sigma}$ equivalent to $L=N_{\sigma}a$. 
          We have produced four complete new sets of data for $\ns=12,18,26$
          and 36 at 74,49,29 and 26 different couplings, respectively. 
          Our updates consisted of
          one heatbath and two overelaxation steps, we measured every fifth
          update. The number of measurements was between 20000 and 80000 at 
          each coupling. 
          We have calculated three observables:
          the Polyakov loop, corresponding to the magnetization, the
          susceptibility in the broken phase $\chi= V (\OPQ - \MO^2)$,
          and that for the symmetric phase $\chi_v = V \OPQ$. Their amplitudes 
          are $B,C_-$ and $C_+$.
\newpage
\section{SCALING ANALYSIS OF THE DATA}

          As input of our analysis we use the same set of critical exponents 
          as \cite{Case97}
          \begin{equation}
          \beta = 0.327~,~~\gamma = 1.239~,~~\nu = 0.631~.
          \end{equation}
          The critical coupling was determined in \cite{Eng96} to 
          $4/g_c^2 = 2.29895(10)$. An analysis
          with our new data fully confirms this result. Moreover it is 
          essentially independent of $\omega$ for $\omega=1.1-1.3$.

\setlength{\unitlength}{1cm}
\begin{picture}(7.5,16)
       \put( 0.8,2.6){
   \epsfig{bbllx=206,bblly=78,bburx=409,bbury=728,
      file=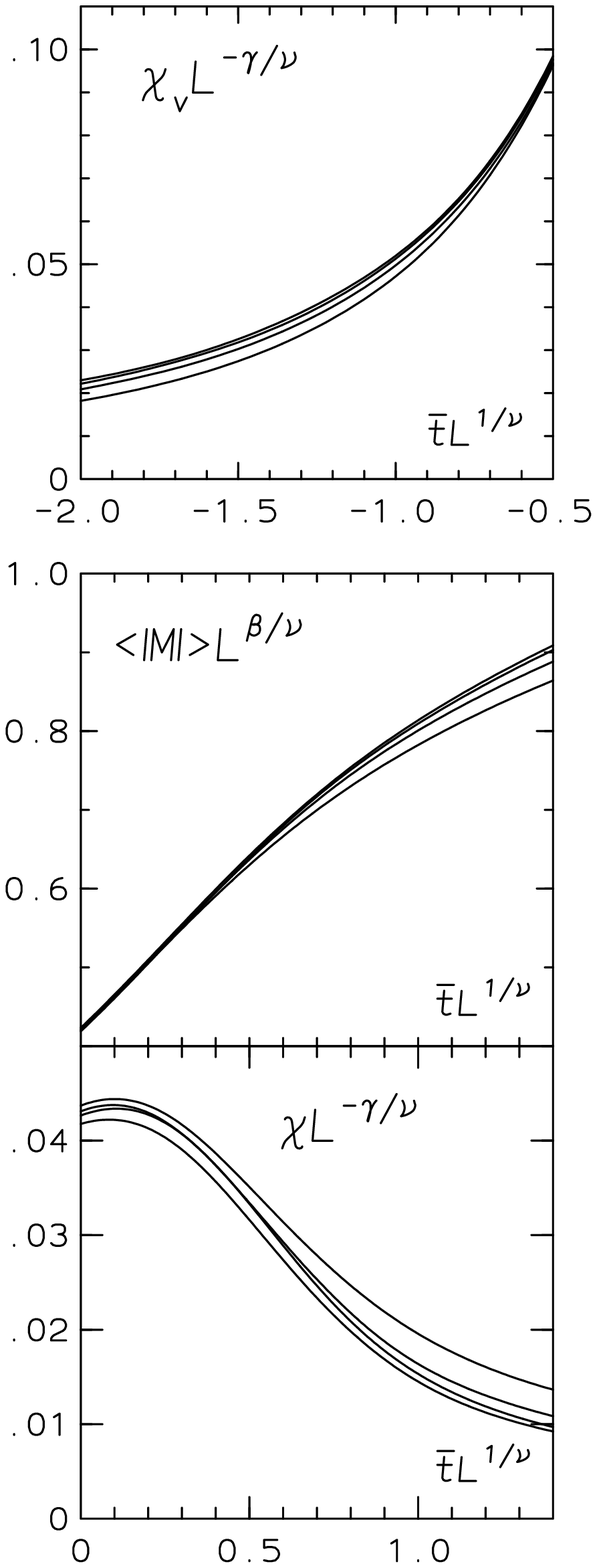, width=50mm,height=130mm}
                    }
       \put( 0.0,0.2){
\begin{minipage}[b]{6.6cm}
        Figure 1. The functions $Q=OL^{-\rho/\nu}$ for
        $L=12,18,26$ and 36. For $Q_M$ and $Q_{\chi_v}$
        the lowest curve is the one for $L=12$, for $Q_{\chi}$
        the order is opposite.
\end{minipage}
                     }
\end{picture}

\noindent In $SU(2)$ the reduced temperature is approximated by
          ${\bar t} = ( 4/g^2 - 4/g^2_c )/ 4/g^2_c$~.
          The symmetric phase occurs at $\bar t<0$, just opposite to 
          magnetic systems. Fig. 1 shows the scaling functions $Q$.
          A consistent succession of curves at fixed $x$ for different $L$
          emerged only
          after using very high statistics and many couplings. There is
          a remarkable difference in the correction-to-scaling behaviours
          in the two phases $\bar t<0$ and $\bar t>0$. In the symmetric 
          phase, here for $\chi_v$, the correction-to-scaling contribution
          is linear in $y$, the sign of $a_1$ is negative 
          (see also \cite{But98}). In the broken 
          phase the correction is certainly not linear for small $L$, both 
          in $Q_M$ and $Q_{\chi}$. Therefore we have estimated $Q_0$ from 
          the two largest lattices. The sign of $a_1(\chi)$ is positive. 
          
          In Fig. 2 we show the functions $F_O$ and $A_0$ for $\omega=1.2$~.
          For large $|x|-$values, where $F_O$ is compatible with zero we
          obtain from  $A_0(x)$
          \begin{eqnarray*}
          ~~~B~  & = & 0.825(1)~, \nonumber\\
          ~~~C_+ & = & 0.0587(8)~, \nonumber\\
          ~~~C_- & = & 0.01243(12)~.\nonumber\\
           C_+/C_-&  = &4.72(11)~.\nonumber
          \end{eqnarray*} 
\noindent 
          Our result for $C_+/C_-$ is in excellent agreement with the $3d$
          Ising model value 4.75(3) of \cite{Case97} and the latest field 
          theoretic value 4.79(10) of \cite{GZJ98}.
          In addition we estimate the ratio for the next-to-leading amplitudes
          of the susceptibility to $~a_{1+}/a_{1-}=-0.37(2)~$.
          Variations of $\omega$ lead
          only to slight changes in the critical amplitudes.
          The correction-to-scaling amplitudes are however more affected.
          Here one should note that we describe the whole correction-to-scaling
          contributions with a single term. Consequently, the value of $\omega$
          is somewhat higher as expected from the relation 
          $\theta=\omega\nu=0.51(3)$.

\setlength{\unitlength}{1cm}
\begin{picture}(7.5,16)
       \put( 0.5,1.6){
   \epsfig{bbllx=206,bblly=78,bburx=409,bbury=728,
      file=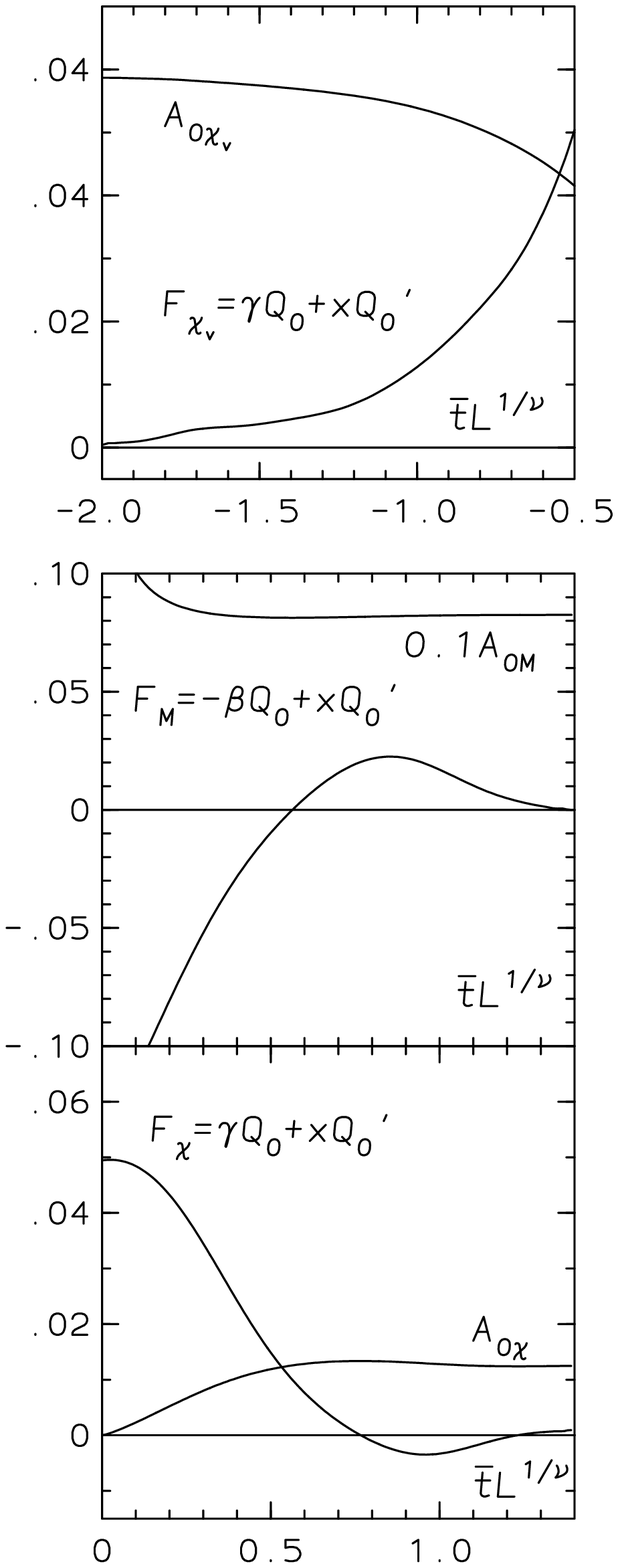, width=56mm,height=143mm}
                    }
       \put( 0.0,0.2){
\begin{minipage}[b]{6.6cm}
        Figure 2. The control functions $F_O$ 
        and the amplitude functions $A_0$ for $\omega=1.2$~.
\end{minipage}
                     }
\end{picture}

\end{document}